\documentclass[sigconf]{acmart}

\usepackage{graphicx}
\usepackage{booktabs}
\usepackage{multirow}
\usepackage{subcaption}
\usepackage{color, colortbl}
\usepackage{arydshln}
\usepackage{balance} 

\definecolor{Gray}{gray}{0.9}

\AtBeginDocument{%
  }

\setcopyright{acmlicensed}
\copyrightyear{2024}
\acmYear{2024}
\acmDOI{XXXXXXX.XXXXXXX}

\acmConference[INRA 2024]{INRA 2024: 12th International Workshop on News Recommendation}{October 14--18, 2024}{Bari, Italy}

\begin{document}

\title{
Peeling Back the Layers: An In-Depth Evaluation of Encoder Architectures in Neural News Recommenders
}

\author{Andreea Iana}
\affiliation{%
  \institution{University of Mannheim}
  \city{Mannheim}
  \country{Germany}
}
\email{andreea.iana@uni-mannheim.de}

\author{Goran Glavaš}
\affiliation{%
  \institution{University of Würzburg}
  \city{Würzburg}
  \country{Germany}}
\email{goran.glavas@uni-wuerzburg.de}

\author{Heiko Paulheim}
\affiliation{%
  \institution{University of Mannheim}
  \city{Mannheim}
  \country{Germany}
}
\email{heiko.paulheim@uni-mannheim.de}

\renewcommand{\shortauthors}{Iana et al.}

\begin{abstract}
Encoder architectures play a pivotal role in neural news recommenders by embedding the semantic and contextual information of news and users. Thus, research has heavily focused on enhancing the representational capabilities of news and user encoders to improve recommender performance. Despite the significant impact of encoder architectures on the quality of news and user representations, existing analyses of encoder designs focus only on the overall downstream recommendation performance. This offers a one-sided assessment of the encoders' similarity, ignoring more nuanced differences in their behavior, and potentially resulting in sub-optimal model selection.
In this work, we perform a comprehensive analysis of encoder architectures in neural news recommender systems. We systematically evaluate the most prominent news and user encoder architectures, focusing on their (i) representational similarity, measured with the Central Kernel Alignment, (ii) overlap of generated recommendation lists, quantified with the Jaccard similarity, and (iii) the overall recommendation performance. Our analysis reveals that the complexity of certain encoding techniques is often empirically unjustified, highlighting the potential for simpler, more efficient architectures. By isolating the effects of individual components, we provide valuable insights for researchers and practitioners to make better informed decisions about encoder selection and avoid unnecessary complexity in the design of news recommenders.
\end{abstract}



\maketitle

\section{Introduction}
\label{sec:intro}

Content-based neural models have become the state of the art in news recommendation. Neural news recommenders (NNRs) typically comprise a news encoder and a user encoder. The news encoder learns semantically meaningful representations of news articles, whereas the user encoder embeds the preferences of users based on their click history \cite{wu2023personalized}. NNRs take the candidate news articles and a user's reading history as input. The relevance of the candidate to the user is determined by comparing, with a scoring function, the latent representations of the two inputs, generated with the corresponding encoders.  
Given the key role of encoders in NNRs, a significant body of research has focused on improving the quality of news encoding and user modeling to improve recommendation performance \cite{karimi2018news,raza2022news,wu2023personalized}. 

On the one hand, ablation studies of recommenders typically analyze individual model components in isolation, neglecting other architecturally comparable model designs \cite{wu2019naml,wu2019nrms,an2019lstur}. At the same time, we see emerging evidence that widely used NNRs exhibit similar performance despite varying model complexities, and that the overall complexity of the recommenders' architecture could be reduced \cite{iana2023simplifying,moller2024explaining}. This highlights the need for a more granular comparison of the individual building blocks to understand their behavior and impact on the overall system. While \citet{moller2022understanding} or \citet{iana2023simplifying} evaluated NNR components such as scoring functions and training objectives, a systematic analysis of encoder architectures is still lacking. Such insights would enable researchers and practitioners alike to make more informed choices about encoder selection in NNR design.
 
On the other hand, progress in the architectural design of news and user encoders is generally measured in terms of the recommender's overall classification and ranking capability \cite{wu2019naml,an2019lstur,qi2022news,wu2021empowering,iana2023simplifying}. Nonetheless, the quality of the embeddings produced by the news and user encoders is equally crucial, given the reliance of the recommender on the dense retrieval paradigm.
Therefore, evaluating NNRs and their components solely in terms of downstream recommendation performance provides a simplified perspective, potentially overlooking subtle differences in the encoders' behavior. We thus argue that investigating the similarity of embeddings generated by various news and user encoders would offer a more nuanced understanding of their behavior, in turn benefiting the model selection process.

In this work, we perform a systematic analysis of the encoder architectures of NNRs. Unlike conventional evaluation studies, we isolate the effects of each core component to the largest possible extent. 
Concretely, we analyze the most prominent news and user encoder architectures in terms of (i) the similarity of learned news, and respectively, user representations, using the Central Kernel Alignment \cite{kornblith2019similarity} metric, (ii) the similarity of the generated recommendation lists, quantified by means of the Jaccard coefficient, and (iii) the impact on the overall recommendation performance. 
Our findings provide a better understanding of news recommenders encoder architectures, not only from a recommendation performance perspective, but also in terms of their representational similarity. We demonstrate that the complexity of some encoding techniques is often empirically unjustified, emphasizing the potential benefits of simpler, more efficient architectures.
These results fundamentally challenge the common practice of over-engineering NNR encoders. Consequently, we derive three key takeaways, arguing that (1) the semantic richness of news encoders is crucial for effective recommendation, that (2) user encoders can be significantly simplified without sacrificing performance, and lastly, (3) we advocate for more rigorous evaluation to guide better informed model selection.

\section{Related Work}
\label{sec:related_work}

Neural news recommenders have significantly advanced in recent years, with encoder architectures playing a key role in capturing the semantic and contextual information of news articles and user profiles. Consequently, a large strand of work has focused on improving the representational capabilities of recommenders by developing ever more accurate, and often complex, news encoding and user modeling architectures. As such, these works have analyzed individual aspects of the NNR components, such as the use of different attention mechanisms in the news or user encoder \cite{wu2019naml,wu2019nrms,qi2020privacy}, the impact of various user modeling \cite{an2019lstur,qi2022news,wang2022news,qi2020privacy,iana2023simplifying} or news embedding \cite{wang2018dkn,wu2019naml,wu2019nrms,qi2020privacy,wu2021empowering,li2022miner,iana2024news} techniques,  or the importance of modeling different news features \cite{wang2018dkn,wu2019naml,wu2019tanr,wu2020sentirec,xun2021we,qi2021pp,wu2021mm} and user characteristics \cite{wu2019nrms,an2019lstur,wu2019nrhub,wu2022feedrec}. Ablation studies in these cases are usually conducted in isolation for the component under consideration, without taking into account the broader architectural context. 

In contrast, another strand of work has started evaluating the impact of NNR components or training strategies across an array of recommendation approaches. For example, \citet{wu2021empowering} have investigated the usage of various pretrained language models as the backbone of widely used NNRs. \citet{moller2022understanding} have evaluated the impact of scoring functions, whereas \citet{iana2023simplifying} have analyzed different user modeling techniques and training objectives. The latter have highlighted the similar recommendation performance achieved by certain models despite differences in architectures and complexity, emphasizing the potential to simplify the design of news recommender systems. While these works shed new light on core components of the recommendation model, their evaluation is most often solely based on the downstream recommendation performance.

The similarity of encoders in NNRs can additionally be measured in terms of their generated representations. More generally, there exist numerous methods for quantifying the similarity of neural networks. Two main categories include (i) representational similarity, which assesses differences in the activations of intermediate layers of neural networks,  and (ii) functional similarity, which compares the networks' outputs in relation to their task \cite{klabunde2023similarity}. Several works have focused on evaluating the representational similarity of (large) language models \cite{wu2020similarity,klabunde2023towards,brown2023understanding,freestone2024word} or of embedding models in Retrieval Augmented Generation systems \cite{caspari2024beyond}, which are often employed as the news encoding component of NNRs.

Nevertheless, to the best of our knowledge, no work so far compares neither user encoders nor news encoders with respect to representational and functional similarity.
In this work, we fill this gap by comprehensively analyzing the primary components of NNR encoder architectures for both news and user inputs.

\begin{table}[t]
    \centering
    \caption{Abbreviations and their description.}
    \label{tab:abbreviations}
    \resizebox{\columnwidth}{!}{%
    \begin{tabular}{ll}
        \toprule
        \textbf{Abbreviation} & \textbf{Description} \\ \midrule
        CNN & convolutional neural network \cite{kim-2014-convolutional}\\
        Att & attention network \\
        AddAtt & additive attention \cite{bahdanau2014neural} \\ 
        MHSA & multi-head self-attention \cite{vaswani2017attention} \\
        PLM & pre-trained language model \\ 
        PLM\textsubscript{[CLS]} & the PLM's output \texttt{[CLS]} token representation \\
        PLM\textsubscript{tokenemb+Att} & PLM's token embeddings pooled with an attention network \cite{wu2021empowering} \\
        SE & sentence encoder \\ \hdashline
        Con & concatenation \\
        Linear & linear layer \\ \hdashline
        LF & late fusion \cite{iana2023simplifying} \\
        GRU & gated recurrent unit \cite{cho2014learning} \\
        CandAware & candidate-aware user encoder \cite{qi2022news} \\
        \bottomrule
    \end{tabular}%
    }
\end{table}

\section{Methodology}
\label{sec:methodology}

We firstly introduce the building blocks of personalized NNRs. Afterwards, we discuss metrics to evaluate both the recommendation performance, as well as the representational similarity of the news and user encoders.

\subsection{Encoders of Neural News Recommenders}

\begin{table*}[t]
    \centering
    \caption{Text encoder architectures.}
    \label{tab:text_encoder}
    \begin{tabular}{lll}
        \toprule
        \textbf{Text Embedding Type} & \textbf{Text Encoder} & \textbf{References} \\ \midrule
        \multirow{3}{*}{word embeddings} 
            & \texttt{CNN + AddAtt} & \cite{wu2019naml,an2019lstur,wu2019tanr,wu2019nrhub,sheu2020context,santosh2020mvl} \\ 
            & \texttt{MHSA + AddAtt} & \cite{gao2018fine,wu2019nrms,wu2020sentirec,wu2020user,ge2020graph,tran2021deep,qi2021pp,wu2022two,wu2021usergraph,wu2021fairness,qi2022news,wang2022news} \\
            & \texttt{CNN + MHSA + AddAtt} & \cite{qi2020privacy} \\ \hdashline
        \multirow{3}{*}{language model} 
            & \texttt{PLM}\textsubscript{\texttt{tokenemb+Att}} & \cite{wu2020nrnf,wu2021empowering,zhang2021amm,zhang2021unbert} \\
            & \texttt{PLM}\textsubscript{[CLS]} & \cite{jia2021rmbert,wu2022news,li2022miner,shivaram2022reducing,iana2023train} \\
            & \texttt{SE} & \cite{iana2024news} \\ 
        \bottomrule
    \end{tabular}%
\end{table*}

Content-based neural news recommenders consists of a dedicated \textbf{(i) news encoder (NE)} and a \textbf{(ii) user encoder (UE)} \cite{wu2023personalized}. The NE transforms different input features (e.g., title, abstract, categories, named entities, images) of a news article $n$ into a latent news representation $\mathbf{n}$. The UE aggregates the embeddings of the clicked news $\mathbf{n}^u_i$ from a user's $u$ history into a user-level representation $\mathbf{u}$. 
Finally, the embedding of a candidate news $\mathbf{n}^c$, outputted by the NE, is scored against the user representation $\mathbf{u}$ produced by the UE, to determine the relevance of the candidate to the user $s(\mathbf{n}^c, \mathbf{u})$. The dot product of the two embeddings $\mathbf{n}^c$ and $\mathbf{u}$ is the most common scoring function \cite{wu2019naml}. NNRs are trained via conventional classification objectives \cite{huang2013learning} with negative sampling \cite{ijcai2022infonce}, or contrastive objectives \cite{iana2023train, liu2023perconet}. 
The building blocks of NNRs (i.e., NE, UE, scoring function, training objective) altogether drive the overall performance of the recommender. 
Since the NE and UE determine what information of the documents and users is embedded by the model, and ultimately, propagated through the recommendation pipeline, both types of encoders play a similarly important role in model selection.  
We introduce the abbreviations used for the remainder of the paper in Table \ref{tab:abbreviations}.

\vspace{1.4mm} \noindent \textbf{News Encoder Architectures.}
The NE can generally be decomposed into a \textit{text encoder}, which embeds the textual content of a news article, and several \textit{feature-specific encoders} (e.g., category, sentiment, entity encoder), which learn to represent further input features different from text chunks. While the former represents a key component of all NNRs, the latter types of encoders are optional and only utilized whenever the textual content is enriched with additional features which might capture or emphasize other aspects of a news article. Lastly, the NE combines the intermediate embeddings produced by the text and feature-specific encoders into a news-level representation by means of a \textit{multi-feature aggregation strategy}. 

We list the most used types of text encoders that we consider in our analysis in Table \ref{tab:text_encoder}, alongside examples of NNRs using them. We distinguish between text encoders that rely on pretrained word embeddings, contextualized by means of convolutional or self-attention networks, and the more recent architectures that employ pretrained language models.\footnote{Note that in this work we do not evaluate encoders which rely on news or user graphs, as such graphs are heavily dataset-dependent. We instead focus on the most used core components of encoders, and leave the analysis of graph-based techniques for future work.}
We additionally consider the most common multi-feature aggregation approaches used to integrate text and other content feature (e.g., category) embeddings into the unified news representation, as shown in Table \ref{tab:multifeat_aggregation}.

\begin{table}[t]
    \centering
    \caption{Multi-feature aggregation strategies for combining textual and categorical representations of news.}
    \label{tab:multifeat_aggregation}
    \begin{tabular}{ll}
        \toprule
        \textbf{Multi-feature aggregation} & \textbf{References} \\ \midrule
        \texttt{AddAtt} & \cite{wu2019naml,santosh2020mvl,wu2021usergraph,sun2021hybrid,raza2021deep,wang2022news} \\
        \texttt{Linear} & \cite{tran2021deep,qi2022news} \\
        \texttt{Con} & \cite{an2019lstur,ge2020graph,han2021neural} \\
        \bottomrule
    \end{tabular}%
\end{table}

\vspace{1.4mm} \noindent \textbf{User Encoder Architectures.}
Parameterized UEs represent the most popular user modeling technique. They learn user representations by means of sequential or attentive networks that contextualize the embeddings of clicked news based on patterns in the user's click behavior. UEs can be further differentiated into candidate-agnostic (i.e., users are encoded separately from candidate news) and candidate-aware (i.e., the user-level aggregation contextualizes the embeddings of clicked news against the embedding of each candidate) encoders \cite{iana2023simplifying}. 
More recently, \citet{iana2023simplifying} proposed the parameter-free late fusion (\texttt{LF}) approach. \texttt{LF} first averages the clicked news embeddings $\mathbf{n}_i^u$ to a user embedding $\frac{1}{N}{\sum^N_{i=1}\mathbf{n}_i^u} = \mathbf{u}$. The inner product of the embedding of the candidate news $\mathbf{n}^c$ and the user embedding $\mathbf{u}$ then represents the relevancy score. Table \ref{tab:user_encoder} lists the main user encoder architectures that we evaluate in this work, together with examples of models using them.

\begin{table}[t]
    \centering
    \caption{User encoder architectures.}
    \label{tab:user_encoder}
    \begin{tabular}{ll}
        \toprule
        \textbf{User Encoder} & \textbf{References} \\ \midrule
        \texttt{LF} & \cite{iana2023train,iana2024news} \\ 
        \texttt{AddAtt} & \cite{gao2018fine,wu2019naml,wu2019nrhub,wu2019tanr,liu2020kred,santosh2020mvl,han2021neural,zhang2021combining} \\
        \texttt{MHSA+AddAtt} & \cite{wu2019nrms,wu2020sentirec,wu2020nrnf,wu2022two,xun2021we,wu2021fairness} \\ \hdashline
        \texttt{GRU}\textsubscript{\texttt{ini}} & \cite{an2019lstur,sun2021hybrid} \\
        \texttt{GRU}\textsubscript{\texttt{con}} & \cite{an2019lstur,tran2021deep}\\
        \texttt{GRU+MHSA+AddAtt} & \cite{qi2020privacy} \\ \hdashline
        \texttt{CandAware} (\texttt{CNN+MHSA+AddAtt}) & \cite{qi2022news}\\
        \bottomrule
    \end{tabular}%
\end{table} 

\subsection{Similarity Evaluation}

We evaluate NEs and UEs on three dimensions: (i) downstream recommendation performance, (ii) similarity of generated recommendations, and (iii) similarity of learned news or user representations. 

\vspace{1.4mm} \noindent \textbf{Downstream Recommendation Performance.}
NNRs are usually evaluated with regards to classification (e.g., AUC) and ranking (e.g., MRR, nDCG) performance. In this work, we focus on the ranking performance, which we quantify using nDCG$@k$. 

\vspace{1.4mm} \noindent \textbf{Similarity of Generated Recommendations.}
We analyze the retrieval similarity of recommenders that use different news or user encoder architectures by the similarity of their top-$k$ recommended articles. Specifically, for the same set of users, we firstly generate the corresponding recommendation lists $R$ and $R'$ with models $M$ and $M'$, respectively. We then measure the similarity of retrieved results with the Jaccard similarity coefficient:

\begin{equation}
    Jaccard(R, R') = \frac{|R \bigcap R'|}{|R \bigcup R'|}
\end{equation}

where $|R \bigcap R'|$ denotes the set of articles recommended by both models, and $|R \bigcup R'|$ the union of all unique news recommended by the two models. The Jaccard similarity score is bounded in the $[0, 1]$ interval, with 1 indicating that both models recommend an identical set of news. Note that the lengths of both recommendation lists will be equal to the full set of candidate news $N^c_u$ for a given user $u$, namely $|R| = |R'| = |N^c_u|$, regardless of the recommendation model used. Thus, to differentiate the retrieval performance of two models, we compute the Jaccard similarity only for the top-$k$ recommendations, ordered descendingly by the recommendation scores. Note that in comparison to nDCG$@k$, the Jaccard similarity measures the overlap of the recommended news between two models without considering the order of the articles in the recommendation set. 

\vspace{1.4mm} \noindent \textbf{Embedding Similarity.}
Numerous measures quantify the representational similarity of neural networks \cite{klabunde2023similarity}. Many of these methods require an identical dimensionality of the compared embeddings or an alignment of the latent representation spaces across models. Since these constraints are not straightforwardly met by the embeddings produced with different news and user encoder architectures, we choose to measure the similarity of embeddings using the Centered Kernel Alignment (CKA) with a linear kernel \cite{kornblith2019similarity}.
Concretely, for a given representation $\mathbf{E}$, we firstly mean-center it column-wise. Afterwards, we compute the pair-wise similarity of the representation of each instance $i$ to all other instances in $\mathbf{E}$. Each row $i$ in the resulting similarity matrix $\mathbf{S}$ thus comprises the similarity between instance's $i$ embedding and all other embeddings, including itself. For two different models with the same number of embeddings $\mathbf{E}$ and $\mathbf{E'}$, the resulting representational similarity matrices $\mathbf{S}$ and $\mathbf{S'}$, respectively, can  be directly compared using the Hilbert-Schmidt Independence Criterion (HSIC) \cite{gretton2005measuring} as follows:

\begin{equation}
    CKA(\mathbf{E}, \mathbf{E'}) = \frac{HSIC(\mathbf{S}, \mathbf{S'})}{\sqrt{HSIC(\mathbf{S}, \mathbf{S}) HSIC(\mathbf{S'}, \mathbf{S'})}}
\end{equation}

The CKA similarity scores are bounded to the interval $[0, 1]$, with a score of 1 denoting equivalent representations. 

\begin{figure*}[t!]
     \centering
     \includegraphics[width=\textwidth]{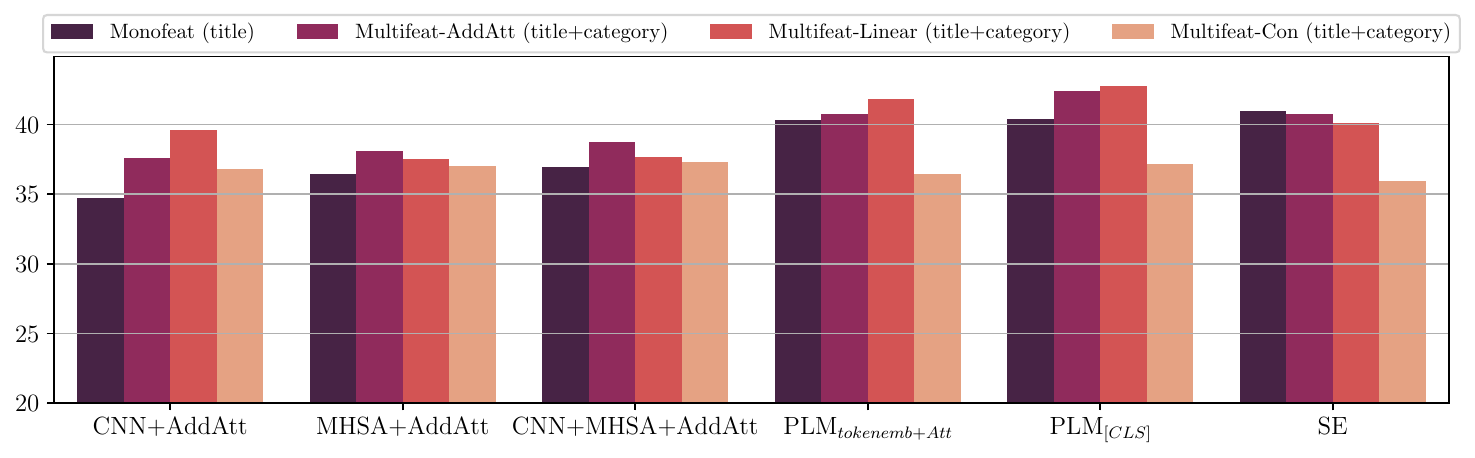}
     \caption{Ranking performance (nDCG@10) of recommenders depending on the news encoder architecture and input features.}
     \label{fig:news_encoder_ndcg}
\end{figure*}

\section{Experimental Setup}
\label{sec:exp_setup}

\vspace{1.4mm} \noindent \textbf{Data.}
We conduct experiments on the MINDsmall \cite{wu2020mind} dataset. Since \citet{wu2020mind} do not release the test set labels, we use the validation portion for testing, and split the respective training set into temporarily disjoint training (the first four days of data) and validation (the last day of data) subsets. 

\vspace{1.4mm} \noindent \textbf{Evaluation Setup.}
We separately evaluate the encoder architectures of NNRs. In all experiments, we consider both mono-feature (e.g., title) and multi-feature (e.g., title and categories) inputs for the NE. In the latter case, we learn category representations by means of a linear encoder that combines a category ID embedding layer with a dense layer \cite{wu2019naml,an2019lstur,qi2022news,wang2022news}. Moreover, in our analysis of NE architectures, we adopt the \textit{late fusion} approach \cite{iana2023simplifying} instead of the traditional parameterized UEs. This evaluation setup allows us to isolate the effects of NEs and to avoid additional confounding factors stemming from the UE, which also influence the output of the NNR. Similarly, when evaluating the similarity of UE architectures, we keep the underlying NE of the recommender fixed, i.e., we analyze different UEs for the same base NE.

\vspace{1.4mm} \noindent \textbf{Implementation and Optimization Details.}
We train all models with the standard cross-entropy loss, using dot product as the scoring function. We use 300-dimensional pretrained Glove embeddings \cite{pennington2014glove} to initialize the word embeddings of the word embedding-based text encoders. Additionally, we use RoBERTa-base \cite{liu2019robertarobustlyoptimizedbert} and the news-specialized multilingual sentence encoder NaSE \cite{iana2024news} for the PLM-based and SE-based text encoders, respectively. We fine-tune only the last four layers of the language models. 
Following prior work \cite{ijcai2022infonce}, we sample four negatives per positive example during training. We set the maximum history length to 50 and train all models with mixed precision, the Adam optimizer \cite{kingma2014adam}, and a batch size of 8. We train all NNRs with word embedding-based NEs for 20 epochs, and those with language model-based NEs for 10 epochs. 
We tune the main hyperparameters of all NNRs using grid search. Concretely, we search for the optimal learning rate in $\{1e{-3}, 1e{-4}, 1e{-5}\}$. We optimize the number of heads in the multi-head self-attention networks in $[8, 12, 16, 20, 24, 32]$, and the query vector dimensionality by sweeping the interval $[50, 200]$ with a step of 50. 
We run all experiments using the implementations available in the NewsRecLib library \cite{iana2023newsreclib}, on a cluster with virtual machines, training each model on a single NVIDIA A100 40GB GPU.\footnote{https://github.com/andreeaiana/newsreclib}

\section{Results and Discussion}
\label{sec:results}

We begin by analyzing the similarity of core NE architectures, followed by an evaluation of UE similarity using the same base news encoding approach. In both cases, we first compare the architectures in terms of ranking performance and retrieval similarity, as these are standard evaluation approaches in the recommender systems field. We then assess the architectures from the perspective of pair-wise embedding similarity.

\subsection{News Encoder Architectures}
\label{sec:ne_results}

\begin{figure}[t]
     \centering
     \includegraphics[width=\columnwidth]{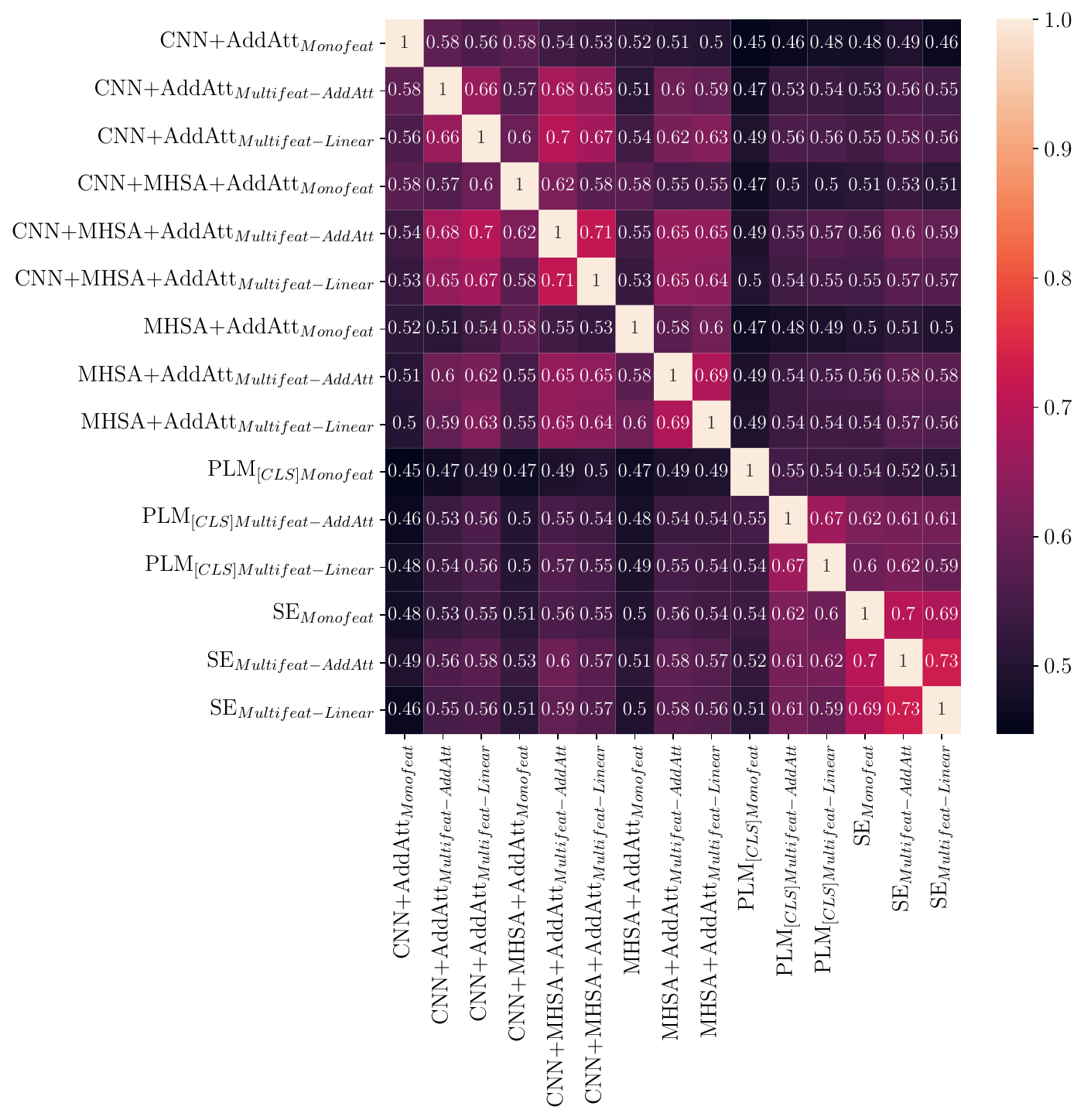}
     \caption{Jaccard similarity for the top-10 recommended news for models with different news encoder architectures and input features. Each model's subscript indicates the type of input, and the multi-feature aggregation strategy, if used.}
     \label{fig:ne_jaccard_10_heatmap}
\end{figure}

\begin{figure}[t]
     \centering
     \includegraphics[width=\columnwidth]{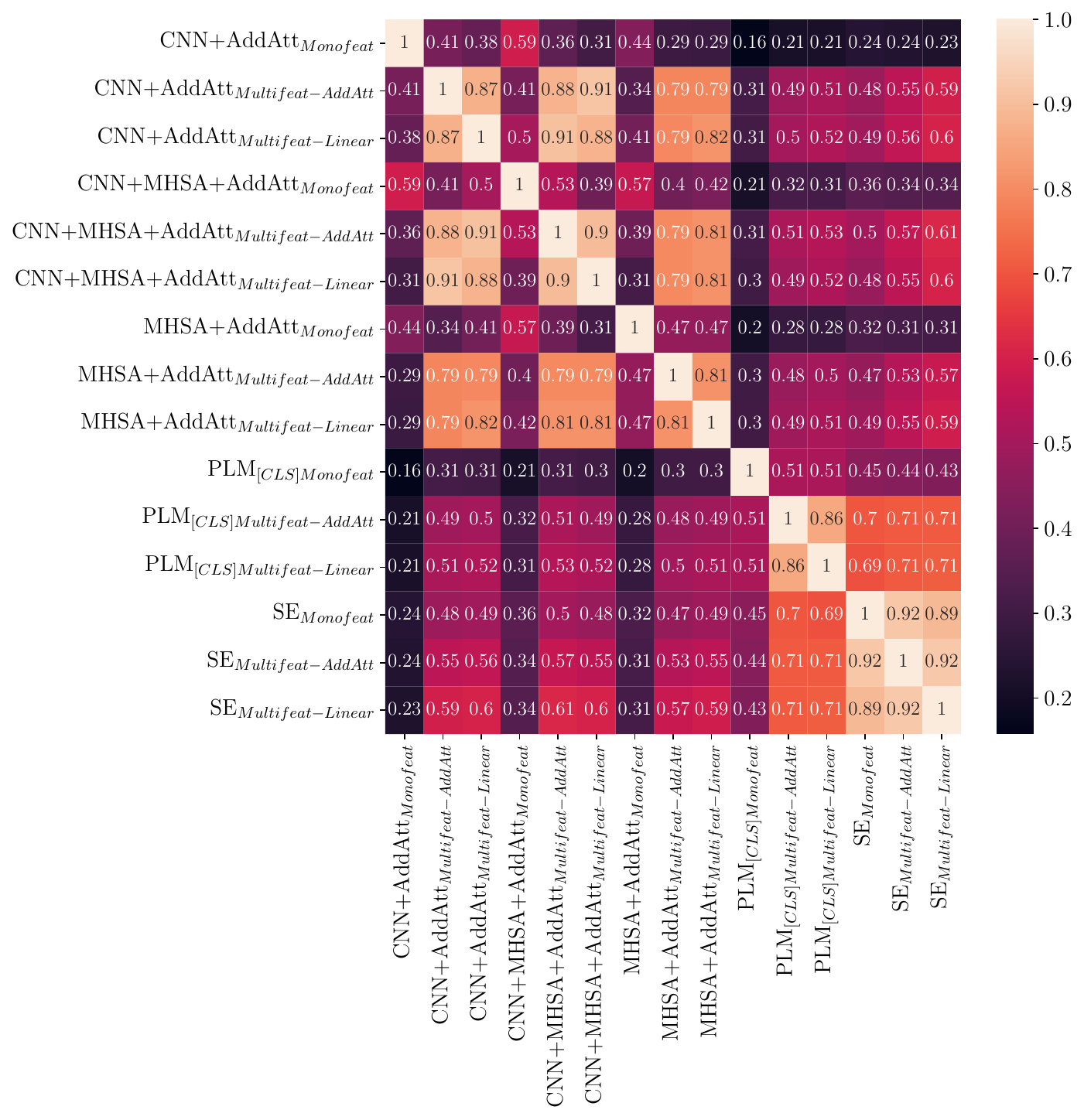}
     \caption{CKA similarity of news embeddings produced with different news encoder architectures and input features. Each model's subscript indicates the type of input, and the multi-feature aggregation strategy, if used.}
     \label{fig:cka_news_heatmap}
\end{figure}

Figure \ref{fig:news_encoder_ndcg} shows the ranking performance, in terms of nDCG@10, of NNRs for different news encoders and input features. For the same input type, e.g. mono-feature, we find a high similarity between the performance of recommenders based on the same family of text encoders. Specifically, text encoders using pretrained static word embeddings are outperformed by those based on PLMs. 
Moreover, \texttt{MHSA+AddAtt} and \texttt{CNN+MHSA+AddAtt} appear to have nearly identical performance, despite the increased complexity of the latter architecture. Similarly, simply using the $[CLS]$ token representation produced by the PLM instead of pooling tokens with an attention network as proposed by \citet{wu2021empowering} leads to slightly better performance while maintaining a lighter text encoder. 

Our findings show that among the three multi-feature aggregation strategies, the \texttt{Linear} and \texttt{AddAtt} approaches always outperform the \texttt{Con} technique. This is intuitive, as the concatenation of vectors with varying dimensionality from non-aligned representation spaces will be sub-optimal. In contrast, both other aggregation strategies project the intermediate text and category embeddings in the same latent representation space. Most importantly, we find that leveraging categories in addition to textual news content as input features is most beneficial for word embedding-based text encoders, and becomes irrelevant or slightly detrimental for the domain-adapted sentence encoder. This can be explained, on the one hand, by the better representational capabilities of the much larger language models which acquire contextual understanding during pretraining compared to static word embeddings. On the other hand, sentence encoders, especially domain-specialized models such as NaSE \cite{iana2024news}, better capture nuances and topics from text due to their pretraining objectives that focus on the overall sentence-level semantics.

\begin{figure*}[t]
     \centering
     \begin{subfigure}[b]{0.46\textwidth}
         \centering
        \includegraphics[width=\columnwidth]{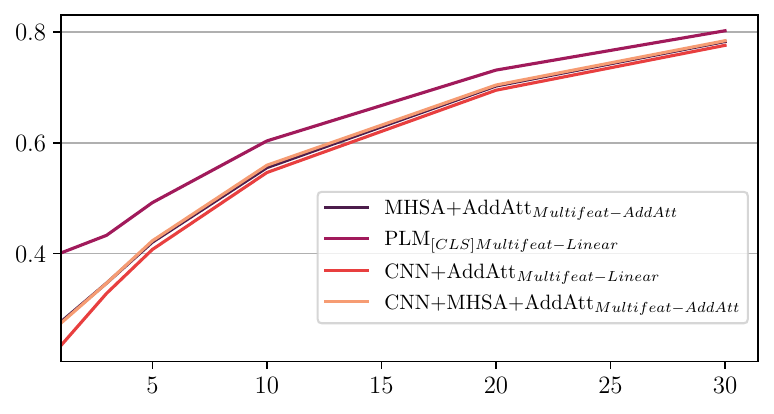}
        \caption{\texttt{SE}\textsubscript{Monofeat} against the best performing architectures from the other news encoder families evaluated.}
        \label{fig:ne_jaccard_se_vs_best_others}
     \end{subfigure}
     \hfill
     \begin{subfigure}[b]{0.46\textwidth}
         \centering
         \includegraphics[width=\columnwidth]{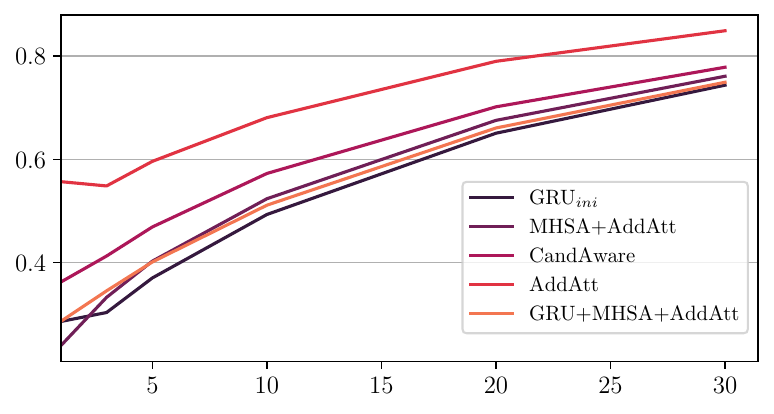}
         \caption{LF against other user encoder architectures evaluated, with \texttt{CNN+AddAtt} as the base news encoder.}
        \label{fig:ue_jaccard_cnnaddatt_lf_vs_best_others}
     \end{subfigure}

    \caption{Evolution of Jaccard similarity for different values of $k$.}
    \label{fig:jaccard}
\end{figure*}

We find these similarities in ranking performance between the various news encoding architectures to be reflected in the similarity of retrieved articles. Figure \ref{fig:ne_jaccard_10_heatmap} illustrates the pair-wise Jaccard similarity scores between the top-10 recommended news per model. Note that we exclude \texttt{PLM}\textsubscript{tokenemb+Att}, as well as the \texttt{Con} multi-feature aggregation strategy from further analysis for the sake of brevity and due to their poorer performance. 
As expected, models from the same family of text encoders show higher similarity scores. The lower Jaccard similarities across word embedding and PLM-based intra-family models using mono-feature versus multi-feature input supports our previous observation regarding the low relevance of categorical input for the domain-adapted SE.

The overall pair-wise Jaccard similarities could initially suggest that most  NEs result in little overlap in their recommendation lists. However, a Jaccard similarity score of 0.54 between two models for a list of $k=10$ recommended items means that, in practice, the two models output 7 identical articles. Analogously, a score of 0.66 indicates an overlap of 8 out of 10 recommendations. As Figure \ref{fig:ne_jaccard_10_heatmap} shows, the recommendations generated by the various NE architectures differ by more than 3 articles in a list of length 10 only in rare cases. In other words, regardless of the architectural differences and complexities, the encoders retrieve, on average, the same articles in over 70\% of the time.

Taking a look at the CKA similarity of the test set news embeddings produced with the different NEs, shown in Figure \ref{fig:cka_news_heatmap}, corroborates our hypothesis: intra-family NEs tend to produce similar embeddings when using the same type of input features. The news-adapted SE constitutes the only exception, as its embeddings are not significantly influenced by leveraging categories as additional input features. Additionally, we observe a higher representational similarity between the \texttt{CNN+AddAtt}, \texttt{MHSA+AddAtt}, and \texttt{CNN+MHSA+AddAtt} models with multi-feature input, and a slightly lower similarity between \texttt{PLM}\textsubscript{[CLS]} and \texttt{SE}-based models. Overall, the high similarity of representations, of recommendation performance, and the large overlap of generated recommendations by the \texttt{CNN+AddAtt}, \texttt{MHSA+AddAtt}, and \texttt{CNN+MHSA+AddAtt} multi-feature NEs contest the empirical contribution of incremental architectural changes in the NE architecture of some NNRs. 

Lastly, we contrast the representational similarity of models against their retrieval similarity. Figure \ref{fig:ne_jaccard_se_vs_best_others} illustrates the evolution of Jaccard similarity scores between the \texttt{SE}\textsubscript{Monofeat} encoder and the best performing architecture from each remaining NE family for different values of $k$. For low values of $k$, we observe a lower similarity of retrieved news for inter-family text encoders, with scores converging toward 1 for larger $k$. An important insight here is that for low values of $k$ (e.g., $k<10$), the news articles retrieved by different NEs tend to be identical, on average, in more than half of the recommended items (e.g., a Jaccard of 0.42 for $k=5$ translates into an overlap of 3 out of 5 items). We observe this behavior even for models with lower representational similarity scores, e.g., word embedding-based NEs versus language model-based NEs. This is relevant from a practical perspective, where retrieval similarity is of most interest for small values of $k$. It would imply, on the one hand, that the representational similarity of NEs might not directly correlate with the retrieval performance for small $k$. On the other hand, this evidence re-affirms our earlier hypothesis that small differences in the architecture and complexity of news encoders do not result in large differences in the actual recommended items.

\subsection{User Encoder Architectures}
\label{sec:ue_results}

\begin{figure*}[t]
     \centering
     \includegraphics[width=\textwidth]{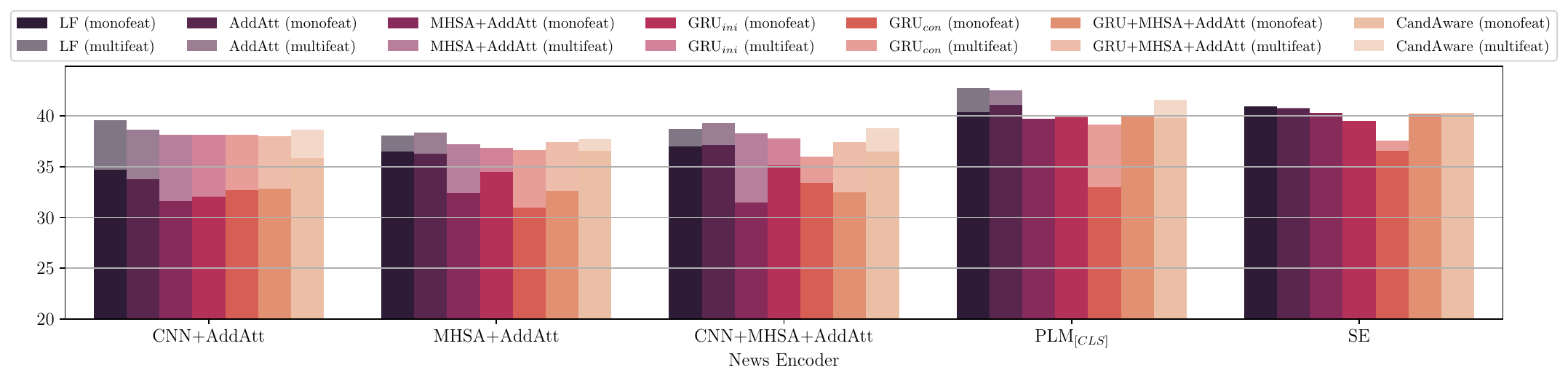}
     \caption{Ranking performance of different recommenders (nDCG@10) depending on the user encoder architecture, for different base news encoder families. The dark bars denote the ranking obtained when using a mono-feature input (i.e., title) in the news encoder, whereas the lighter bars indicate the (generally higher) scores gained with a multi-feature input (i.e., title and category), and the best multi-feature aggregation strategy per news encoder family.}
     \label{fig:user_encoder_ndcg}
\end{figure*}

\begin{figure}[t]
     \centering
     \includegraphics[width=\columnwidth]{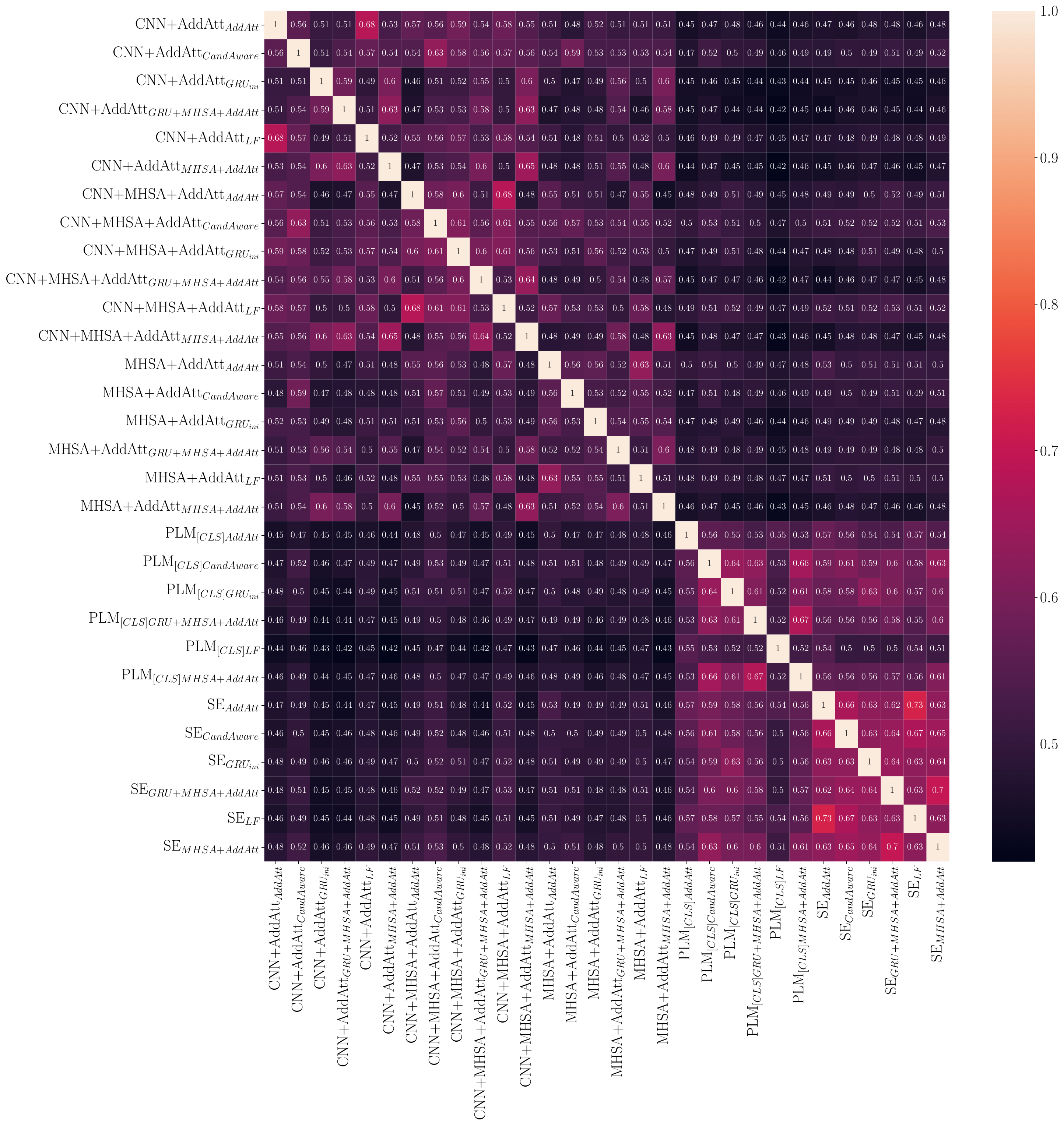}
     \caption{Jaccard similarity for the top-10 recommended news for models with different user encoder architectures. Each model name denotes the base news encoder, with the user encoder architecture indicated by the subscript.}
     \label{fig:ue_jaccard_10_heatmap}
\end{figure}

\begin{figure}[t]
     \centering
     \includegraphics[width=\columnwidth]{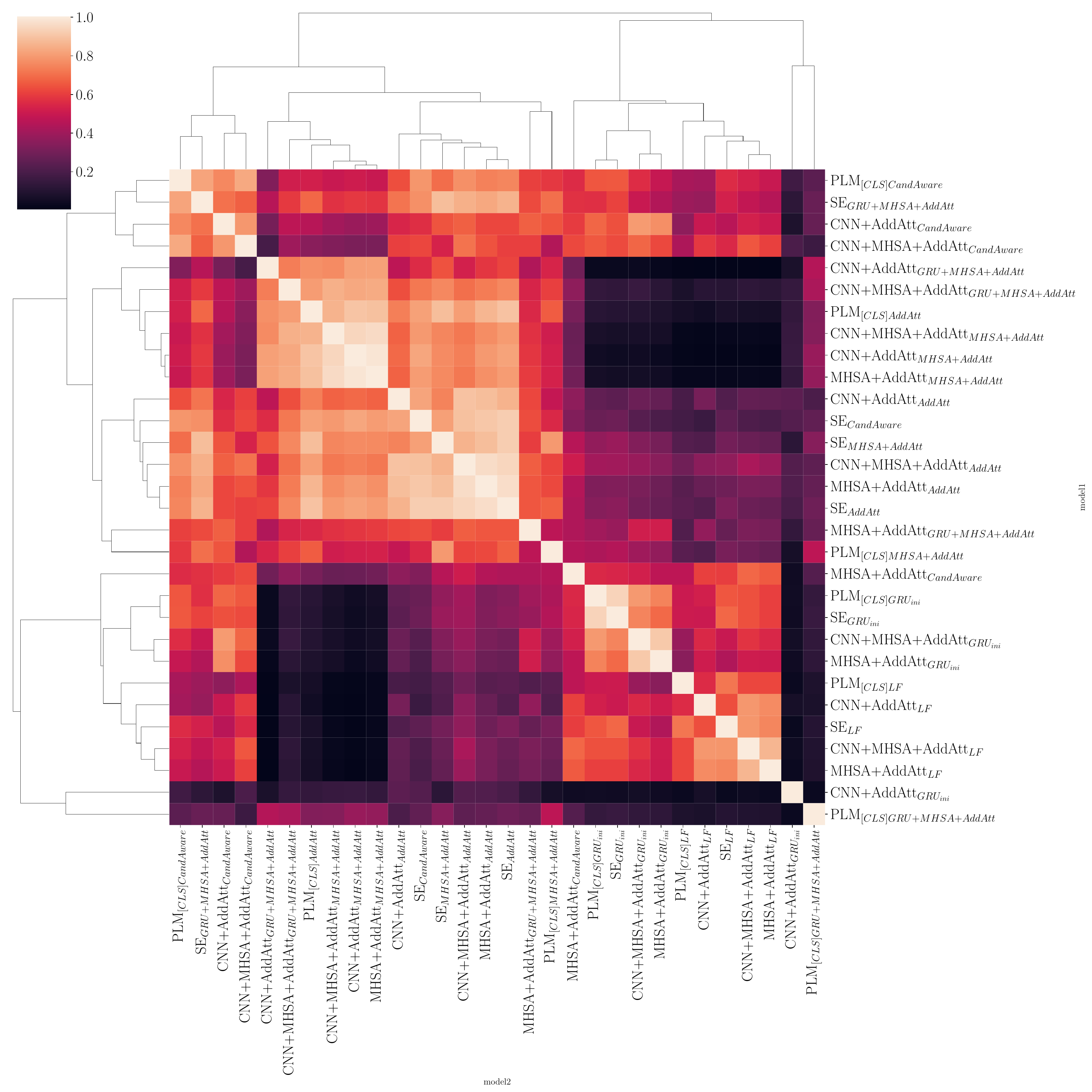}
     \caption{CKA similarity of user embeddings produced with different user encoders, for different families of base news encoders. Each model name denotes the base news encoder, with the user encoder architecture indicated by the subscript.}
     \label{fig:cka_user_monofeat_clustermap}
\end{figure}

We next investigate the ranking performance, with regards to nDCG@10, for different UE architectures for the same base NE. Figure \ref{fig:user_encoder_ndcg} displays the corresponding results, for both mono-feature, and well as multi-feature input. We find that the \texttt{LF}, \texttt{AddAtt}, and \texttt{CandAware} encoders perform the best across all families of NEs. More specifically, the much simpler \texttt{LF} and \texttt{AddAtt} encoders outperform the complex \texttt{CandAware} modeling technique in the case of language model-based NEs, and perform similarly with \texttt{CandAware} for word embedding-based NEs, as previously suggested by \citet{iana2023simplifying}. 
Surprisingly, these two approaches also consistently achieve better ranking than sequential-based UEs (i.e., \texttt{GRU+MHSA+AddAtt}, \texttt{GRU}\textsubscript{ini}, \texttt{GRU}\textsubscript{con}). Once again, we see that using categorical information alongside the textual content as input to the NE benefits all recommenders regardless of the UE family. The only exception, as previously discussed, are SE-based NNRs. Interestingly, we see that multi-feature inputs close the gap (i) in between inter-family UEs for the same base NE, and (ii) across intra-family UEs for different underlying NEs. 
Most importantly, our findings corroborate earlier results from \citet{iana2023simplifying} and \citet{moller2024explaining} that the complexity of user encoders can be simplified, particularly when the bi-encoder NNR leverages language models pretrained, or even domain-specialized, on large-scale corpora, to obtain news representations.

The heatmap in Figure \ref{fig:ue_jaccard_10_heatmap} shows the Jaccard similarity scores for the top-10 recommendations, for the different UE families, when using only the title as input to the NE.\footnote{The results with multi-feature input are similar, and we omit them for the sake of brevity.} We exclude \texttt{GRU}\textsubscript{con} from further analysis as it underperforms the counterpart variant \texttt{GRU}\textsubscript{ini}. 
We observe that in terms of retrieval similarity, the NNRs are clustered based on the underlying NE family, regardless of the UE used. Once again, the results indicate a large overlap of recommended news (i.e., on average, of at least 7 out of 10 recommendations) for the UEs within these clusters. Moreover, we observe comparable similarity patterns across inter-family UEs for the same NE family; different NEs change only the absolute magnitude of the Jaccard similarity scores. Within intra-family clusters of NEs, the findings re-affirm that \texttt{LF} and \texttt{AddAtt} have the highest overlap in terms of the top-10 recommended articles; their generated recommendations usually differ in at most 2 or 3 items, on average. This is intuitive, as \texttt{LF} represents a special case of \texttt{AddAtt}, where the attention weights are all equal, and set to the inverse of the user's history length. 

We delve deeper into the retrieval similarity of UE architectures. Figure \ref{fig:ue_jaccard_cnnaddatt_lf_vs_best_others} shows the Jaccard similarity of \texttt{LF} against the other user modeling approaches for a recommender with a \texttt{CNN+AddAtt}-based NE, for different values of $k$. As in Section \ref{sec:ne_results}, the Jaccard similarity of recommended news is sensitive to the value of $k$, with scores converging toward 1 for larger values of $k$. On the one hand, the scores of sequential UEs (\texttt{GRU}\textsubscript{ini}, \texttt{GRU+MHSA+AddAtt}) are clustered closely together, which can be explained by their shared sequential component. However, the retrieved articles appear to be more similar between sequential and non-sequential UEs (e.g., higher Jaccard similarity between \texttt{GRU+MHSA+AddAtt} and \texttt{MHSA+AddAtt}) across intra-family NEs, than between sequential UEs. This could be attributed to the architectural differences of the two models, among which \texttt{GRU+MHSA+AddAtt} employs an attention network similar to that of \texttt{MHSA+AddAtt}. These mixed results, combined with the better performing non-sequential UEs, call into question the efficiency of modeling the news recommendation task as a sequential recommendation problem \cite{wu2022news}.

We shift our attention to the pair-wise similarity of user embeddings generated by the different types of UEs for the users in the test set, illustrated in the heatmap of Figure \ref{fig:cka_user_monofeat_clustermap}. We additionally perform a hierarchical clustering on the heatmap to identify clusters of similar UEs \cite{waskom2021seaborn}. In contrast to retrieval results, we find that the architecturally comparable families of UEs dictate the similarity of embeddings, regardless of the underlying NE used. Most surprisingly, we find that although the top-recommended news by \texttt{GRU}\textsubscript{ini} and \texttt{GRU+MHSA+AddAtt} moderately overlap, their user representations are highly dissimilar. Moreover, the latent representations of \texttt{AddAtt} appear more similar to other attention-based UEs than with \texttt{LF}. This could be explained by the fact that as a particular case of \texttt{AddAtt}, the parameterless \texttt{LF} does not reshape the embedding space, as it simply computes an average of the user's clicked news. Nonetheless, these differences in the representational similarities of UEs also do not appear to directly correlate with more dissimilar retrieval performance. This suggests that in real-world applications, the lightweight and conceptually simple \texttt{LF} constitutes an equally effective and more efficient alternative to \texttt{AddAtt}, and especially, to more complex architectures.

\subsection{Key Takeaways}
Following the results of our in-depth analysis of the embedding and retrieval similarity of the most prominent news and user encoder architectures, we highlight several key takeaways. 

\vspace{1.4mm} \noindent \textbf{Semantic Richness is Key.}
Our analysis demonstrates that the semantic richness of news encoders, achieved either through multi-feature input or contextualized language models, significantly outweighs the impact of UEs. This is particularly the case when initializing news representations with large-scale PLMs. Additionally, contextualized language models can effectively capture semantic nuances, such as topical information, without heavily relying on categorical annotations. From a practical standpoint, this reduces the need for manual or automatic feature engineering, streamlining the NNR design process. We hence argue that research on news encoding should focus more on leveraging and adapting existing semantically informed, contextualized language models for the task of news recommendation, rather than on incrementally modifying existing architectures.

\vspace{1.4mm} \noindent \textbf{User Encoders Can be Considerably Simplified.}
Our findings show that retrieval similarity is primarily influenced by the underlying NE family, rather than the specific UE used. At the same time, simpler approaches such as \texttt{LF} and \texttt{AddAtt} not only result in significantly better ranked results, but their retrieved items largely overlap with those recommended by more complex UE architectures. These findings thus render simpler architectures as better and more lightweight user modeling alternatives. Additionally, the high retrieval similarity between parameter-free (i.e., \texttt{LF}) and parameterized (e.g., \texttt{AddAtt}) encoders heavily indicates that, in practice, there is little empirical justification for an additional parameterized component in the news recommender system. Furthermore, the similarity of sequential and non-sequential encoders indicates that treating news recommendation as a sequential problem might be sub-optimal. We speculate that the high item churn characteristic of news, combined with short user histories, limit the benefits of differentiating between long and short-term user preferences, in contrast to other domains, such as movie or book recommendation. In conclusion, in line with \citet{moller2024explaining}, we posit that user modeling should not focus exclusively on the architectural component, but instead, should pay closer attention to the users' motivations to consume certain news, on the one hand, and to collecting richer and more accurate user (relevance) feedback, on the other hand.

\vspace{1.4mm} \noindent \textbf{More Rigorous Evaluation is Needed for Better Model Selection.}
Our findings, along with recent research \cite{moller2022understanding,iana2023simplifying,moller2024explaining},  highlight the limitations of current evaluation practices in news recommendation. By focusing solely on performance metrics, we risk overlooking critical aspects of model behavior, leading to sub-optimal component selection and incremental model advancement. Therefore, we advocate for a more comprehensive and rigorous evaluation approach. Ablation studies should consider the broader architectural context, and together with model comparisons, should extend beyond performance-based evaluation to include a more granular behavioral and representational analysis. This would provide a more nuanced understanding of model similarities and differences, guiding researchers and practitioners toward better informed model selection decisions.

\section{Conclusion}
\label{sec:conclusion}

Despite the central role played by encoder architectures in neural news recommenders, their advancement and understanding is generally limited to one-sided evaluation in terms of recommendation performance.
In this work, we conducted a comprehensive evaluation of encoder architectures in neural news recommenders, by systematically analyzing their (i) representation similarity, (ii) overlap of generated recommendations, and (iii) overall recommendation performance. 
Evaluations of recommenders on standard benchmarks often reveal insignificant performance differences between compared models or among their ablated components. Consequently, our analysis of differences in representational similarity and retrieval overlap of neural news recommenders serves as a complementary evaluation tool for understanding the relationship between the architectural design, behavior, and downstream performance of models.  

Our findings offer more nuanced insights into the interplay of news and user encoders, and challenge the assumption that complex encoding techniques are essential for accurate news recommendation.  We demonstrate that simpler, yet equally effective architectures can yield comparable results. This underscores the importance of understanding recommenders' behavior from multiple perspectives, and of balancing model complexity with performance.
Specifically, we emphasize three key takeaways: (1) the crucial role of semantic richness in news encoders, (2) the potential for simplifying user encoders without sacrificing accuracy, and (3) the need for more rigorous evaluation and ablation studies to inform architectural design choices.
By fostering a more transparent and nuanced understanding of encoder architectures in neural news recommenders, we hope to guide researchers and practitioners toward more efficient and effective model designs.
\begin{acks}
The authors acknowledge support by the state of Baden-Württemberg through bwHPC and the German Research Foundation (DFG) through grant INST 35/1597-1 FUGG. We also thank Fabian David Schmidt for proof-reading.
\end{acks}

\bibliographystyle{ACM-Reference-Format}
\balance
\bibliography{references}

\end{document}